\documentstyle[12pt,epsfig]{article}

\catcode`\@=11
\def\@citex[#1]#2{%
\if@filesw \immediate \write \@auxout {\string \citation {#2}}\fi
\@tempcntb\m@ne \let\@h@ld\relax \def\@citea{}%
\@cite{%
  \@for \@citeb:=#2\do {%
    \@ifundefined {b@\@citeb}%
      {\@h@ld\@citea\@tempcntb\m@ne{\bf ?}%
      \@warning {Citation `\@citeb ' on page \thepage \space undefined}}%
      {\@tempcnta\@tempcntb \advance\@tempcnta\@ne%
      \@tempcntb\number\csname b@\@citeb \endcsname \relax%
      \ifnum\@tempcnta=\@tempcntb 
        \ifx\@h@ld\relax%
          \edef \@h@ld{\@citea\csname b@\@citeb\endcsname}%
        \else%
          \edef\@h@ld{\ifmmode{-}\else--\fi\csname b@\@citeb\endcsname}%
        \fi%
      \else
        \@h@ld\@citea\csname b@\@citeb \endcsname%
        \let\@h@ld\relax%
      \fi}%
    \def\@citea{,\penalty\@highpenalty\,}%
  }\@h@ld
}{#1}}

\def\@citeb#1#2{{[#1]\if@tempswa , #2\fi}}
\def\@citeu#1#2{{$^{#1}$\if@tempswa , #2\fi }}
\def\@citep#1#2{{#1\if@tempswa , #2\fi}}

\def\bcites{         
        \catcode`\@=11
        \let\@cite=\@citeb
        \catcode`\@=12
}

\def\upcites{         
        \catcode`\@=11
        \let\@cite=\@citeu
        \catcode`\@=12
}

\def\plaincites{      
        \catcode`\@=11
        \let\@cite=\@citep
        \catcode`\@=12
}

\newcount\hour
\newcount\minute
\newtoks\amorpm
\hour=\time\divide\hour by 60
\minute=\time{\multiply\hour by 60 \global\advance\minute by-\hour}
\edef\standardtime{{\ifnum\hour<12 \global\amorpm={am}%
        \else\global\amorpm={pm}\advance\hour by-12 \fi
        \ifnum\hour=0 \hour=12 \fi
        \number\hour:\ifnum\minute<10 0\fi\number\minute\the\amorpm}}
\edef\militarytime{\number\hour:\ifnum\minute<10 0\fi\number\minute}

\def\draftlabel#1{{\@bsphack\if@filesw {\let\thepage\relax
   \xdef\@gtempa{\write\@auxout{\string
      \newlabel{#1}{{\@currentlabel}{\thepage}}}}}\@gtempa
   \if@nobreak \ifvmode\nobreak\fi\fi\fi\@esphack}
        \gdef\@eqnlabel{#1}}
\def\@eqnlabel{}
\def\@vacuum{}
\def\marginnote#1{}
\def\draftmarginnote#1{\marginpar{\raggedright\scriptsize\tt#1}}
\def\draft{
        \pagestyle{plain}
        \overfullrule=2pt
        \oddsidemargin -.5truein
        \def\@oddhead{\sl \phantom{\today\quad\militarytime} \hfil
        \smash{\Large\sl DRAFT} \hfil \today\quad\militarytime}
        \let\@evenhead\@oddhead
        \let\label=\draftlabel
        \let\marginnote=\draftmarginnote
        \def\ps@empty{\let\@mkboth\@gobbletwo
        \def\@oddfoot{\hfil \smash{\Large\sl DRAFT} \hfil}
        \let\@evenfoot\@oddhead}
        \def\@eqnnum{(\theequation)\rlap{\kern\marginparsep\tt\@eqnlabel}%
        \global\let\@eqnlabel\@vacuum}  }

\def\blackfonts{
        \font\blackboard=msbm10 scaled\magstep1
        \font\blackboards=msbm8
        \font\blackboardss=msbm6
}

\def\nblack{            
        \def\ZZ{{Z \n{10} Z}}
        \def\NN{{N \n{14} N}}
        \def\CC{{C \n{11} C}}
        \def\RR{{R \n{11} R}}
        \def\QQ{{Q \n{12} Q}}
}

\def\prep{         
        \catcode`\@=11
        \input art10.sty
        \catcode`\@=12
        
        \let\small\null
        \def\blackfonts{
                \font\blackboard=msbm10
                \font\blackboards=msbm7
                \font\blackboardss=msbm5
        }
        \let\sl\it
        \twocolumn
        \sloppy
        \voffset=-2.54truecm
        \hoffset=-2.54truecm
        \flushbottom
        \parindent 1em
        \leftmargini 2em
        \leftmarginv .5em
        \leftmarginvi .5em
        \marginparwidth 48pt
        \marginparsep 10pt
        \setlength{\columnsep}{2truecm}
        \setlength{\textwidth}{25.4truecm}
        \setlength{\textheight}{17truecm}
        \baselineskip=16pt
        \oddsidemargin .18truein
        \evensidemargin .17truein
}

\def\eqalign#1{\null\,\vcenter{\openup\jot\m@th
  \ialign{\strut\hfil$\displaystyle{##}$&$\displaystyle{{}##}$\hfil
      \crcr#1\crcr}}\,}
\def\eqalignno#1{\displ@y \tabskip\centering
  \halign to\displaywidth{\hfil$\@lign\displaystyle{##}$\tabskip\z@skip
    &$\@lign\displaystyle{{}##}$\hfil\tabskip\centering
    &\llap{$\@lign##$}\tabskip\z@skip\crcr
    #1\crcr}}

\def\section{\@startsection {section}{1}{\z@}{3.ex plus 1ex minus
 .2ex}{2.ex plus .2ex}{\large\bf}}
\def\subsection{\@startsection{subsection}{2}{\z@}{2.75ex plus 1ex minus
 .2ex}{1.5ex plus .2ex}{\bf}}        

\def\appendix{{\newpage\section*{Appendix}}\let\appendix\section%
        {\setcounter{section}{0}
        \gdef\thesection{\Alph{section}}}\section}

\def\abstract{\if@twocolumn
\section*{Abstract}
\else 
\begin{center}
{\bf Abstract\vspace{-.5em}\vspace{0pt}}
\end{center}
\quotation
\fi}

\catcode`\@=12

\newcommand{\wh}{\widehat}

\newcommand{\hp}{{\wh\Phi}}
\newcommand{\hq}{{\wh Q_B}}
\newcommand{\he}{{\wh\eta_0}}
\newcommand{\ha}{{\wh{A}}}

\newcommand{\lllb}{\Bigl\langle\Bigl\langle}
\newcommand{\rrrb}{\Bigr\rangle\Bigr\rangle}
\newcommand{\beq}{\begin{equation}}
\newcommand{\eeq}{\end{equation}}
\newcommand{\beqa}{\begin{eqnarray}}
\newcommand{\eeqa}{\end{eqnarray}}

\def\noj#1,#2,{{\bf #1} (19#2)\ }
\def\jou#1,#2,#3,{{\sl #1\/ }{\bf #2} (19#3)\ }
\def\ann#1,#2,{{\sl Ann.\ Physics\/ }{\bf #1} (19#2)\ }
\def\cmp#1,#2,{{\sl Comm.\ Math.\ Phys.\/ }{\bf #1} (19#2)\ }
\def\ma#1,#2,{{\sl Math.\ Ann.\/ }{\bf #1} (19#2)\ }
\def\ng#1,#2,{{\sl Nagoya.\ Math.\ J.\/ }{\bf #1} (19#2)\ }
\def\jd#1,#2,{{\sl J.\ Diff.\ Geom.\/ }{\bf #1} (19#2)\ }
\def\invm#1,#2,{{\sl Invent.\ Math.\/ }{\bf #1} (19#2)\ }
\def\cq#1,#2,{{\sl Class.\ Quantum Grav.\/ }{\bf #1} (19#2)\ }
\def\cqg#1,#2,{{\sl Class.\ Quantum Grav.\/ }{\bf #1} (19#2)\ }
\def\ijmp#1,#2,{{\sl Int.\ J.\ Mod.\ Phys.\/ }{\bf A#1} (19#2)\ }
\def\jmphy#1,#2,{{\sl J.\ Geom.\ Phys.\/ }{\bf #1} (19#2)\ }
\def\jams#1,#2,{{\sl J.\ Amer.\ Math.\ Soc.\/ }{\bf #1} (19#2)\ }
\def\grg#1,#2,{{\sl Gen.\ Rel.\ Grav.\/ }{\bf #1} (19#2)\ }
\def\mpl#1,#2,{{\sl Mod.\ Phys.\ Lett.\/ }{\bf A#1} (19#2)\ }
\def\nc#1,#2,{{\sl Nuovo Cim.\/ }{\bf #1} (19#2)\ }
\def\np#1,#2,{{\sl Nucl.\ Phys.\/ }{\bf B#1} (19#2)\ }
\def\pl#1,#2,{{\sl Phys.\ Lett.\/ }{\bf #1B} (19#2)\ }
\def\pla#1,#2,{{\sl Phys.\ Lett.\/ }{\bf #1A} (19#2)\ }
\def\pr#1,#2,{{\sl Phys.\ Rev.\/ }{\bf #1} (19#2)\ }
\def\prd#1,#2,{{\sl Phys.\ Rev.\/ }{\bf D#1} (19#2)\ }
\def\prl#1,#2,{{\sl Phys.\ Rev.\ Lett.\/ }{\bf #1} (19#2)\ }
\def\prp#1,#2,{{\sl Phys.\ Rept.\/ }{\bf #1C} (19#2)\ }
\def\ptp#1,#2,{{\sl Prog.\ Theor.\ Phys.\/ }{\bf #1} (19#2)\ }
\def\ptpsup#1,#2,{{\sl Prog.\ Theor.\ Phys.\/ Suppl.\/ }{\bf #1} (19#2)\ }
\def\rmp#1,#2,{{\sl Rev.\ Mod.\ Phys.\/ }{\bf #1} (19#2)\ }
\def\yadfiz#1,#2,#3[#4,#5]{{\sl Yad.\ Fiz.\/ }{\bf #1} (19#2) #3%
\ [{\sl Sov.\ J.\ Nucl.\ Phys.\/ }{\bf #4} (19#2) #5]}
\def\zh#1,#2,#3[#4,#5]{{\sl Zh.\ Exp.\ Theor.\ Fiz.\/ }{\bf #1} (19#2) #3%
\ [{\sl Sov.\ Phys.\ JETP\/ }{\bf #4} (19#2) #5]}

\def\beq{\begin{equation}}
\def\eeq{\end{equation}}
\def\beqar{\begin{eqnarray}}
\def\eeqar{\end{eqnarray}}

\newcommand{\be}{\begin{equation}}
\newcommand{\ee}{\end{equation}}
\newcommand{\bea}{\begin{eqnarray}}
\newcommand{\eea}{\end{eqnarray}}

\def\nfrac#1#2{{\displaystyle{\vphantom1\smash{\lower.5ex\hbox{\small$#1$}}%
        \over\vphantom1\smash{\raise.25ex\hbox{\small$#2$}}}}}

\def\n#1{\mskip-#1mu}

\def\lae{\mathrel{\mathop{\smash{\lower .5 ex \hbox{$\stackrel<\sim$}}}}}
\def\lae{\mathrel{\mathop{\smash{\lower .5 ex \hbox{$\stackrel>\sim$}}}}}

\def\bbbz{{\sf Z\!\!\!Z}}

\def\Tr{{\rm Tr}}
\def\l:{\mathopen{:}\,}
\def\r:{\,\mathclose{:}}

\def\fracs#1#2{\textstyle\frac #1#2}

\catcode`\@=11
\def\theequation{\arabic{equation}}
\catcode`\@=12

\nblack
\bcites

\nblack

\catcode`\@=11
\def\theequation{\thesection.\arabic{equation}}
\@addtoreset{equation}{section}
\@addtoreset{footnote}{section}
\@addtoreset{footnote}{subsection}
\catcode`\@=12

\newcommand{\beqn}{\begin{equation}}
\newcommand{\eeqn}{\end{equation}}
\newcommand{\beqnarray}{\begin{eqnarray}}
\newcommand{\eeqnarray}{\end{eqnarray}}

\newcommand{\rrr}{\rangle \rangle}
\newcommand{\lll}{\langle \langle}
\newcommand{\PSbox}[3]{\mbox{\rule{0in}{#3}
\includegraphics{#1}\hspace{#2}}}
\newcommand {\bear} [1] {\begin {array} {#1}}
\newcommand {\ear} {\end {array}}

\newcommand {\beqarn} {\begin{eqnarray*}}
\newcommand {\eeqarn} {\end{eqnarray*}}

\newcommand{\de}{\partial}

\newcommand {\nn} {\nonumber}

\begin{document}

\begin{titlepage}

\begin{center}
\today
\hfill CTP-MIT-3014\\
\hfill                  hep-th/yymmnn

\vskip 1.5 cm
{\large \bf On Marginal Deformations in Superstring Field Theory}
\vskip 1 cm 
{Amer Iqbal$^{\sharp}$, Asad Naqvi$^{\dagger}$}\\
\vskip 0.5cm

\footnote{ Address from $1^{\mbox{st}}$ September 2000\,:~$^{\sharp}$Department of Physics-Theory Group, The University of Texas at Austin, Austin, TX, 78712.
\\ $^{\dagger}$ Department of Physics and Astronomy, University of 
Pennsylvania, Philadelphia, Pennsylvannia, 19104}{\sl Center for Theoretical Physics, MIT\\
Cambridge, MA 02139, U.S.A.\\}

\end{center}

\vskip 0.5 cm
\begin{abstract}
We use level truncated superstring field
 theory to obtain the
effective potential for the Wilson line marginal deformation
parameter which corresponds to the constant vacuum expectation value
of
 the $U(1)$ gauge field on the D-brane in a particular
direction. We present 
 results for both the
 BPS and the non-BPS
D-brane. In the case of non-BPS D-brane the
 effective potential has
branches corresponding to the extrema of the
 tachyon potential. In
the branch with vanishing tachyon vev ($M$-branch), 
 the effective
potential becomes flatter as the level of the approximation
 is
increased. The branch which corresponds to the stable vacuum after
the tachyon has condensed ($V$-branch) exists only for a finite range
of values of marginal deformation parameter. We use our
results 
 to find the mass of the gauge field in the stable tachyonic vacuum. We
find
 this mass to be of a non-zero value which  seems to stabilize as the level
approximation
 is improved. 
\end{abstract}
\end{titlepage}

\thispagestyle{empty}

{\scriptsize

\tableofcontents}

\pagenumbering{arabic}

\vspace{1cm}

\section{Introduction}
Recently, it has been realized that string field theory is a useful
setup to address questions related to tachyon condensation in string
theory \cite{9911116}. Tachyonic systems, such as the open string
tachyon on a D-brane in bosonic string theory as well as on a non-BPS
D-brane or a D-brane anti D-brane pair in Type II string theory, have been
analyzed in level truncated string field theory
\cite{9912249,0001084,0002211,0002237,0003220,0004015,0004112,0006240}. 
It has also been argued that a co-dimension one D-brane  can be
considered as the tachyonic lump in bosonic open string field theory 
\cite{0002117,0003031,0005036}. Co-dimension two lump solutions in string field theory 
have also
been explored in a recent papers \cite{0008053,0008101}. Tachyon condensation in the D0/D4 
 system in the
presence of NS B-field has also been studied, using the superstring
field theory, in
\cite{0007235}. 
From the two dimensional conformal field
 theory point of view, vevs
for the tachyonic field corresponds to
 adding relevant operators to
the CFT \cite{0003101}. In addition to
 the tachyon, D-branes have
massless open string states some of which
 correspond to exactly
marginal operators of the CFT. Such massless
 states, which correspond
to marginal deformations, should correspond
 to flat directions in the
potential of string field theory.
 
 One familiar example of a
marginal deformation of the CFT is the
 Wilson-line, a constant vev of
a $U(1)$ gauge field along a compact
 direction.  In string field
theory, we expect a flat potential for the
 field $w$ corresponding to
this marginal deformation.  However, in the
 level truncation scheme,
to any finite level, the potential is not
 expected to be exactly
flat. In \cite{0007153} Sen and Zwiebach calculated the potential, in
the level truncated bosonic string field theory, for the marginal
deformation parameter corresponding to the constant vev of the $U(1)$
gauge field. They determined the effective potential for $w$ by
solving the equations of motion for all fields except $w$ in terms of
$w$ and substituting in the potential.  They showed that the effective
potential becomes flatter as the higher level field and terms in the
action are included.  The effective potential, as expected, is not
flat at any given level. However, as the approximation is improved,
the potential becomes flatter.
 
 In section 2, we
 calculate the
effective potential for the Wilson line on a BPS
 D-brane in Type II
string theory by using 
 Berkovits' open
 superstring field theory
\cite{9912121,0001035}. We work in the level
 truncation scheme and
calculate the potential first at the
 (0,0) level and then at the
(1,2) level. By $(n,m)$ level approximation, we mean that we keep the
states up to level $n$ and terms in the action up to level $m$. For
the case of the BPS D-brane, we assign level to a state to be equal
to its $L_0$ eigenvalue. This is different from the level assignments
in section 3 as well as the assignments in
\cite{0002211,0003220,0004015}. At the level (1,2), we did not find 
different branches of solutions in  the BPS case. 

In section 3, we analyze the case of the non-BPS D-brane.  Here, the
tachyon, in the GSO$(-)$ sector survives and has $L_0$ eigenvalue
$-\frac{1}{2}$. We assign level of a state with $L_{0}$ eigenvalue
$h$
 to be $h+\frac{1}{2}$ so that the tachyon has level 0.  A major
difference between the BPS and the non-BPS
 case is the existence of
two distinct branches of solutions in the non-BPS case. One branch
corresponds to zero vev of the tachyon field,$t$, ($M$-branch
\cite{0007153}) and the other branch  corresponds to the
true vacuum after the tachyon has condensed ($V$-branch). That there is a $t=0$ branch for 
all values of $w$ is clear form the fact that GSO odd fields appear at least 
quadratically in the action hence a solution exists to the equations of motion
with all GSO($-$) fields set to zero. Hence the effective potential for $w$
in the $M$-branch is identical to that in the BPS case.   
However, in the non-BPS case, in addition to the solution with $t=0$, 
 there are two
solutions with non-zero vev for the  tachyon but they differ from each other
by a sign. This continues to be the case at non-zero values
of $w$. This is different from the case of bosonic string field 
theory where there was no branch in which the tachyon vev was
identically zero. The two branches in the bosonic case where not
related to each other by an overall sign and were physically
distinct. For the non-BPS D-brane, there are really three branches 
but two of them are trivially related to each other by $t \rightarrow -t$.  
As the value of $w$ increases, the
$V$-branch merges with the $M$-branch at a particular value of
$w$. For values of $w$ greater than this value, the $V$-branch does
not exist. However, the $M$-branch continues to exist for all values
of $w$ (at least in the level approximation up to level $(3/2,3)$). In
section 4
 we find the mass of the gauge
 field in the $V$-branch and
as suggested by Taylor
\cite{0008033,0001201}, we find a non-zero value which seems to stabilize
as the level approximation is improved. 

\section{Marginal deformations and Level Approximation on a BPS D-brane}  
In this section, we will analyze the string fields and the string field 
action corresponding to CFT marginal deformations. In particular, we will
single out a particular coordinate, $x_9$ along the world volume of a D-9
brane, and consider giving expectation value to the constant mode of the
gauge field component $A_{9}$. This represents a marginal deformation of
the BCFT describing the D-brane. We will find the string field corresponding
to this deformation in the level approximation. For our purposes, whether
or not the $x_9$ direction is compact is irrelevant; we give expectation
values to the field independent in the $x_9$ direction, so that 
modes carrying non-zero momentum along the $x_9$ direction (and other
directions) are set to zero. 
We will use Berkovits' open superstring field theory for our analysis. 

\subsection{String field theory on a BPS D-brane} 
In \cite{9503099,9912121,0001084}, a  string field
configuration in the GSO$(+)$ NS sector corresponds to 
a Grassmann even open string vertex operator
$\Phi$ of ghost and picture number 0 \cite{FMS} in the combined
conformal field theory of a $c=15$ superconformal matter system, and
the $b,c,\beta,\gamma$ ghost system with $c=-15$. Here, $b$, $c$ are
the reparameterization ghosts and $\beta$, $\gamma$ are their superconformal
partners. The $\beta$, $\gamma$ ghost system can be bosonized and can 
be replaced by ghost fields $\xi,\eta,\phi$, related to $\beta,\gamma$ through
the relations 
\[ 
\beta=\de\xi e^{-\phi}, \qquad \gamma =\eta e^\phi.
\]
The SL(2,R) invariant vacuum has zero ghost and picture number.

We shall denote by $\langle \prod_i A_i \rangle$ the correlation function
of a set of vertex operators in the combined matter-ghost conformal field
theory on the unit disk
with open string vertex operators inserted on the boundary of the disk.
These
correlation functions are to be computed with the
normalization
\begin{equation} \label{eb1}
\langle \xi(z) c\partial c\partial^2 c(w) e^{-2\phi(y)}\rangle = 2\, .
\end{equation}
The BRST operator is given by 
\be
\label{BRST}
Q_B=\oint dz   c \bigl( T_m + T_{\xi\eta} + T_\phi)
+ c \partial c b\,+\,\eta \,e^\phi
\, G_m\,-\,\eta\partial \eta e^{2\phi} b \,.
\ee
Here \footnote{We have set $\alpha^\prime=2$}
\be
T_{\xi\eta}=\partial\xi\,\eta, \,\, T_\phi=-\fracs{{1}}{{2}} \partial\phi \partial
\phi -\partial^{2}\phi \, ,\,\,
T_m=-\frac{1}{2}\partial X^\mu \partial X_\mu - \, \frac{1}{2}\,\psi^\mu \partial \psi_\mu\,,
\,\, G_m=i \,\psi^\mu \partial X_{\mu}\,. 
\ee
The
normalization of
$\psi^\mu$, $\de X^\mu$, $\phi$, $\xi$, $\eta$, $b$ and $c$ are as follows:
\[\de X^\mu(z) \, \de X^\nu(w) = -\frac{ \eta^{\mu \nu}}{(z-w)^2} \,\,\,\,\,\,\,\,
\psi^\mu(z) \, \psi^\nu(w) = \frac{\eta^{\mu \nu}}{z-w} \]
\[
\xi(z) \eta(w) \simeq {1 \over z-w}, \quad b(z) c(w)\simeq {1\over z-w}, 
\quad \partial \phi(z)
\partial\phi(w)\simeq
-{1\over (z-w)^2} \, .
\]
We will denote by $\eta_0=\oint dz \eta(z)$ the zero mode of 
the field $\eta$ acting on the Hilbert space of matter ghost CFT.

The expansion of the string field action, relevant for our calculations is
given by \cite{0002211}
\bea \label{e17}
S \hskip-5pt&&\hskip-10pt = {1\over g^2} \lllb 
\,\,{1\over 2} 
 \,(Q_B \Phi)\,\,(\eta_0 \Phi)\, 
 + {1\over 3}
\,(Q_B\Phi)\,  \Phi \,(\eta_0\Phi)  +
{1\over 12}  \,(Q_B\Phi) \, \Bigl( \Phi^2\, 
(\eta_0\Phi) - \Phi\,(\eta_0\Phi)\,\Phi \, \Bigr) 
 \cr  
&& \qquad +{1\over 60}
 \,(Q_B\Phi) \, \Bigl( \Phi^3\, (\eta_0\Phi) - 3\Phi^2\,(\eta_0\Phi)\,\Phi \Bigr) 
\cr
&& \qquad  + {1\over 360}\,(Q_B\Phi) \, 
\Bigl( \Phi^4\, (\eta_0\Phi) - 4\Phi^3\,(\eta_0\Phi)\,\Phi
+ 3\Phi^2\,(\eta_0\Phi)\,\Phi^2  \Bigr) 
\rrrb \,. 
\eea
$\lll A_1, \ldots A_n \rrr$
is defined as:
\be \label{e2ff}
\lll A_1\ldots A_n \rrr = \Bigl\langle f^{(n)}_1 \circ A_1(0)\cdots 
f^{(n)}_n\circ A_n(0)\Bigr\rangle\, .
\ee
Here, 
$f\circ A$ for
any function $f(z)$, denotes the conformal transform of $A$ by $f$, and 
\be \label{e3}
f^{(n)}_k(z) = e^{2\pi i (k-1)\over n} \Big({1+iz\over 1-iz}
\Big)^{2/n}\,\quad  \hbox{for} \quad n\geq 1 .   
\ee
\subsection{String fields for the Wilson line marginal deformation}
To study the Wilson line marginal deformations in string field theory, 
 we will restrict the string field to be in 
 the subset of vertex operators of ghost and 
picture number zero, created from the matter stress tensor ($T_m(z)$), 
its superconformal partner ($G_m(z)$), $\psi^9(z)$ and $\de X^9(z)$ (where
$x_9$ is the compact direction) and the ghost fields $b,c,\xi, \eta,
\phi$. From now on, we will refer to $\psi^9$ and $\de X^9$ as just
$\psi$ and $\de X$.  In this section, the level of a string field 
component multiplying a vertex operator will be defined as the
conformal weight $h$  of the vertex operator 
\footnote{This is different from the definition of level in the
next section and also from the definition of level in \cite{0002211,0003220,0004015}}. 
The string field action has a gauge invariance which can be used
to choose a gauge in which 
\be
\label{egauge} 
b_0\Phi=0, \qquad \xi_0\Phi=0\, ,
\ee
which is valid at least at the linearized level.
 All relevant states
in the ``small'' Hilbert space
 can be obtained by acting with
 ghost
number zero combinations of
 oscillators
 $\psi_{-r}$, $\alpha_{-n}$,
$L_n$, $G_{-1-r}$,
 $b_{-n}$, $c_{1-n}$, $\beta_{-r}$, $\gamma_{-r}$
($n \geq 1 $ and $r \geq \frac{1}{2}$) on $| \Omega\rangle$.
 $|
\Omega\rangle$ is the tachyon state (projected out by the GSO
projection).  The $b_0\Phi=0$ gauge condition allows us to 
 ignore
states with a $c_0$ oscillator in them (except when $L_0$ eigenvalue
of the state is zero).  
 As shown in \cite{9705038} and used in
\cite{0007153}, the string field theory action has a $Z_2$ symmetry
which is a combination of twist symmetry and $X\rightarrow -X$ and
$\psi \rightarrow -\psi$. Under twist symmetry, string field
components associated with a vertex operator of dimension $h$ carry
charge $(-1)^{h+1}$ for even $2h$, and $(-1)^{h+{1\over2}}$ for odd
$2h$. Thus it is possible to consistently restrict the string field
$\Phi$ to be even under this $Z_2$ symmetry. We will now work in the
level truncation scheme.  $(n,m)$ approximation means that we keep
states up to level $n$ and terms in the action up to level $m$. States
up to level one which can contribute to the potential for the marginal
deformation parameter are shown in Table 1. The boxed states are odd
under the $\bbbz_{2}$ symmetry mentioned above and therefore can be ignored.

\begin{table}[ht]
\begin{eqnarray*} 
\begin{array}{||c|c|c|c||} \hline
\rule{0mm}{5mm}L_{0}    & \mbox{Level}  & \mbox{Twist}     & \mbox{State}    \\ \hline \hline
\rule{0mm}{8mm}~~0      & 0             & \mbox{odd}       & \fbox{$c_{0}\beta_{-\frac{1}{2}}|\widetilde{\Omega}\rangle$}\,,\,\psi_{-\frac{1}{2}}|\tilde{\Omega}\rangle \\ \hline
\rule{0mm}{8mm}~~1      & 1             & \mbox{even}       &\fbox{$\psi_{-\frac{1}{2}}\beta_{-\frac{1}{2}}\gamma_{-\frac{1}{2}}|\widetilde{\Omega}\rangle$}\,,\,\fbox{$\psi_{-\frac{3}{2}}|\widetilde{\Omega}\rangle$}\,,\,   c_{-1}\beta_{-\frac{1}{2}}|\widetilde{\Omega}\rangle \\
& & & b_{-1}\gamma_{-\frac{1}{2}}|\widetilde{\Omega}\rangle\,\,,
 \,\,G_{-\frac{3}{2}}|\widetilde{\Omega}\rangle \,\,,\,\,\alpha_{-1}\psi_{-\frac{1}{2}}|\widetilde{\Omega}\rangle \\ \hline
\end{array}
\end{eqnarray*} 
\caption{GSO even states of ghost number zero up to level one. Boxed states are
odd under the $\bbbz_{2}$ symmetry mentioned before.}
\end{table}

\subsubsection{Level (0,0)}
Since only GSO(+) sector survives on a BPS D-brane, at $L_0=0$, we can 
construct the state
\bea
\psi_{-\fracs{{1}}{{2}}}| \Omega\rangle \,.
\eea
Notice that the state $c_0 \beta_{-\frac{1}{2}}|\Omega\rangle$ has $L_0$ eigenvalue zero. However, 
it is 
is odd under the $Z_2$ combination of twist and parity mentioned
above so we can ignore it in our present calculation. 

The state $\psi_{-\fracs{{1}}{{2}}}| \Omega\rangle$ corresponds to the 
level 0 gauge field zero mode. The string field at this level is given by
\bea
\Phi^{(0)}= w \, W\,,
\eea
where 
\bea
W\,=\, \xi\, \psi \,c \,e^{-\phi},
\eea
is a primary field of weight zero. 
The string field action can be used to obtain the potential 
\bea
V(\Phi)=-S(\Phi)=-(2\pi^2 M)\,g^2S(\Phi)\,.
\eea
 Here $M=\frac{1}{2\pi^2 g^2}$: 
\bea
V^{(0,0)}_{eff}(w)=(2\pi^2 M)\frac{3}{8}\,w^4
\eea
We note the absence of quadratic term for $w$. This is to be expected since
$W$ is a marginal operator. 
\subsubsection{Level (1,2)}
At $L_0=1$ (level 1), we get four new states \footnote{We have not 
included states $\psi_{-\frac{1}{2}} \beta_{-\frac{1}{2}} \gamma_{-\frac{1}{2}} | \Omega\rangle
$ and $\psi_{-\frac{3}{2}} | \Omega\rangle$ because they are odd under the combination
of twist and parity symmetry.}:
\bea
c_{-1} \beta_{-\frac{1}{2}} | \Omega\rangle, \,\,\,
b_{-1} \gamma_{-\frac{1}{2}} | \Omega\rangle, \,\,\,
G_{-\frac{3}{2}} | \Omega\rangle, \,\,\,
 \psi_{-\frac{1}{2}} \alpha_{-1} | \Omega\rangle.
\eea
The vertex operators corresponding to these states are:
\bea
A=\xi \partial \xi c \partial^{2}c e^{-2\phi}\,,\,E =\xi \eta \,,\,
F=\xi G_{m}ce^{-\phi}\,,\,
P=i\,\xi \psi \partial X ce^{-\phi}\,,
\eea
where $F$ and $P$ are primary fields of conformal weight $1$ and  $A$ and $E$ are not primary but 
transform as ($f(0)=u$) \cite{0002211}:
\bea
f\circ A(0)&=&f'(0)\,\{ A(u)-\fracs{{f''(0)}}{{(f'(0))^{2}}}\,c\partial c\,\xi\,\partial\xi e^{-2\phi}(u)\,\}\\ \nn
f\circ E(0)&=&f'(0)\,\{E(u)-\fracs{{f''(0)}}{{2(f'(0))^{2}}} \,\}\\ \nn 
\eea
The  string field $\Phi$ is given by
\bea
\Phi^{(1)}=\, w\,W\,+\,a \, A \,+\, f \, F\,+ \, p\, P\, + \, e\, E\,. 
\eea
We can use this expansion in the string field action (\ref{e17}) and truncate
the action to terms up to level 2. 

The result, with $S_k$ denoting the level $k$ terms in the action, is
\begin{eqnarray}
g^2 S_0 &=& - \frac{{3}}{{8}} \,w^{4} \\ \nn
g^2 S_1 & = &  \frac{{4}}{{9}} \,\sqrt{3}\,(w^{2}\,a - w^{2}\,f - w^{2}\,p ) + 
\frac{{1}}{{75}}\sqrt{250+110\sqrt{5}}\,w^{4}\,e \,
\,  \\\nn
g^2 S_2 & = & 
- \frac{{1}}{{2}} \,p^{2}  -   \,f\,p 
 - 2 \, a\,e - 5\, \,f^{2} 
   - \frac{{83}}{{24}} \,w^{2}\,a\,e + 
\frac{{95}}{{24}} \,w^{2}\,e\,f  
  - \frac{{39}}{{8}} \, w^{2}\, f^{2} \\ \nn &&  - 
 \frac{{15}}{{16}} \,w^{2}\,p^{2} + \frac{{19}}{{48}} \,w^{2}\,e\,p 
  -  \frac{{15}}{{8}} \,w^{2}\,f\,p + \frac{{19}}{{48}} \,w^{2}\,e^{2}- \frac{{25}}{{177}} \,w^{4}e^2 .
\end{eqnarray}
\bea
V(w,a,e,f,p)=-(2\pi^2 M)\,g^2(S_0+S_1+S_2).
\eea
The equations of motion, $\frac{\de V}{\de a}=\frac{\de V}{\de e}=\frac{\de V}{\de f}=\frac{\de V}{\de p}=0$ for
$a,e,f,p$ can be solved to obtain $a,e,f,p$ as functions of $w$. These can be substituted in $V(w,a,e,f,p)$ to
obtain the effective potential for $w$,
\bea
V^{(1,2)}_{eff}(w)=(2\pi^2 M) \, \frac{P(w)}{Q(w)}
\eea
where
\begin{eqnarray*}
P(w)&=& \frac {w^4}{106200} \Bigl(  29344814625\,w^{8} - 131631360\,\sqrt{3}\,\sqrt{250 + 110\,\sqrt{5}}\,w^{8} + 68413522000\,w^{6} \\
 & & - 296763392\,\sqrt{250 + 110\,\sqrt{5}}\,w^{6}\,\sqrt{3} - 207831040\,\sqrt{3}\,\sqrt{250 + 110\,\sqrt{5}}\,w^{4}\\ & &  + 55895956800\,w^{4} 
  + 17269913600\,w^{2} - 46399488\,\sqrt{3}\,\sqrt{250 + 110\,\sqrt{5}}\,w^{2}\\ &&  + 1232486400 \Bigr)\\ Q(w) & = & 
 147456 + 915456\,w^{2} + 2085184\,w^{4} + 2049104\,w^{6} + 723345\,w^{8} 
\end{eqnarray*}
It is useful to expand
$V^{(1,2)}_{eff}(w)$ in powers of $w$. The result is
\begin{eqnarray}
\frac{V^{(1,2)}_{eff}(w)}{M}& =&  {\displaystyle \frac {17}{216}} \,w^{4} - ( - 
{\displaystyle \frac {199}{324}}  + {\displaystyle \frac {2}{675}
} \,\sqrt{3}\,\sqrt{250 + 110\,\sqrt{5}})\,w^{6}  + 
O(w^{8})
\end{eqnarray}
\begin{figure}[h]
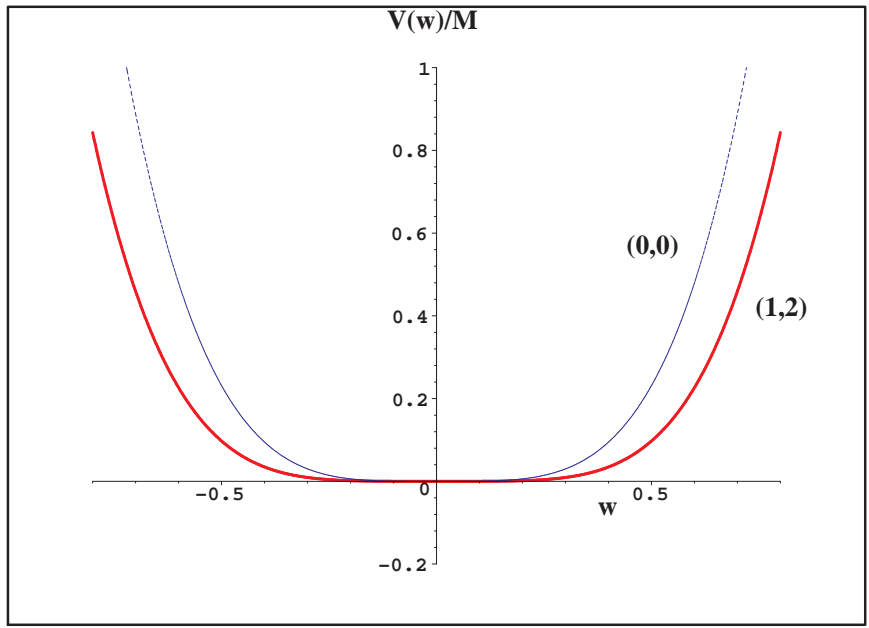

\PSbox{figbps.ps hoffset=30 voffset=280 hscale=50 vscale=50 angle=-90}{4.8in}{3.4in}
\caption{The effective potential for $w$ in the $(0,0)$ and $(1,2)$ approximations for
a BPS D-brane.} 
\end{figure}
The coefficient of the $w^4$ term is $\frac{17}{216}$ in the $(1,2)$ approximation
as compared to $\frac{3}{8}$ in the $(0,0)$ approximation. The potential
becomes flatter as we improve the approximation as shown in Fig 1. However, even at the (1,2)
level, we did not find multiple branches of solutions or any restriction
on the value of $w$. This is different for the case of a D-brane in 
bosonic string theory discussed in \cite{0007153}. 

\section{Marginal deformations and Level approximation on a non-BPS D-brane}
\subsection{String field theory on a non-BPS D-brane}
For a non-BPS D-brane, in addition to the GSO($+$) sector, GSO($-$) 
sector survives. 
 The GSO($-$) states are
Grassmann odd and to incorporate them in the algebraic structure of
the
 previous section, internal $2 \times 2$ Chan Paton matrices were
introduced in \cite{0002211}. These are added both to the vertex
operators, $Q_B$ and $\eta_0$ as discussed in detail in
\cite{0002211}.
 The complete string field, $Q_{B}$ and $\eta_{0}$ are
thus written as
\bea
\widehat{\Phi} = \Phi_{+} \otimes I  + \Phi_- \otimes \sigma_{1}\,,\,\,\,\,
\widehat{Q}_{B}= Q_{B}\otimes \sigma_{3}\,,\,\,\,\,
\widehat{\eta}_{0}=\eta_{0}\otimes \sigma_{3}\,.
\eea
where the subscripts on $\Phi$ denote the $(-1)^{F}$ eigenvalue of the
vertex operator. The string field action for the non-BPS D-brane up to terms quintic 
in the string field is given by \cite{0001084}:
\bea \label{actionsusy}
S \hskip-5pt&&\hskip-10pt = {1\over 2g^2} \lllb 
\,\,{1\over 2} 
 \,(\hq \hp)\,\,(\he \hp)\, 
 + {1\over 3}
\,(\hq\hp)\,  \hp \,(\he\hp)  +
{1\over 12}  \,(\hq\hp) \, \Bigl( \hp^2\, 
(\he\hp) - \hp\,(\he\hp)\,\hp \, \Bigr) 
 \cr  
&& \qquad +{1\over 60}
 \,(\hq\hp) \, \Bigl( \hp^3\, (\he\hp) - 3\hp^2\,(\he\hp)\,\hp \Bigr) 
\cr
&& \qquad  + {1\over 360}\,(\hq\hp) \, 
\Bigl( \hp^4\, (\he\hp) - 4\hp^3\,(\he\hp)\,\hp
+ 3\hp^2\,(\he\hp)\,\hp^2  \Bigr) 
\rrrb \,. 
\eea
where the
correlation function appearing in the above action are defined as
\bea \nn
\lll \wh A_1\ldots \wh A_n \rrr = {\Tr} \Bigl\langle f^{(n)}_1 \circ
\ha_1(0)\cdots
f^{(n)}_n\circ \ha_n(0)\Bigr\rangle \, \,.
\eea
 \subsection{Marginal deformations and Level approximation}
In this section we will calculate the potential for the Wilson line
corresponding to the constant gauge field in the $x^{9}$
direction. The corresponding massless state in the CFT is
$\psi_{-\frac{1}{2}}|\widetilde{\Omega}\rangle$. We will denote the
corresponding space-time field by $w$. As in the last section $\psi:=
\psi^{9}$ and $\partial X:=\partial X^{9}$.

Since we are on a non-BPS D-brane we also have fields in the GSO$(-)$
sector, especially the tachyon, which contribute to the potential for
$w$. In the last section we defined the level of the fields to be
equal to the $L_{0}$ eigenvalue of the corresponding CFT state. In
this section we will have to modify the definition of the level due to
the presence of the tachyon. A field corresponding to the CFT state of
$L_{0}$ eigenvalue $h$ has level $h+\frac{1}{2}$.

In our calculation of the potential for the Wilson line we will
restrict ourselves to fields of level less than or equal to
$\frac{3}{2}$ and terms of level 3 or less in the action.  The states
and
 the vertex operators contributing to the potential are shown in
the Table 2.
\begin{table}[ht]
\begin{eqnarray*} 
\begin{array}{||c|c|c|c|c||} \hline
\rule{0mm}{5mm}L_{0}          & \mbox{Level}  & \mbox{Twist}     & GSO(+)       &GSO(-)    \\ \hline \hline
\rule{0mm}{7mm} -\frac{1}{2} &                 0     &\mbox{even}   &                  &   |\tilde{\Omega}\rangle      \\  \hline
\rule{0mm}{8mm}~~0                & \frac{1}{2}      &\mbox{odd}               & \fbox{$c_{0}\beta_{-\frac{1}{2}}|\widetilde{\Omega}\rangle$}\,,\,\psi_{-\frac{1}{2}}|\tilde{\Omega}\rangle
 &  \\ \hline
\rule{0mm}{7mm} ~~ \frac{1}{2} & 1 &\mbox{even}& &\alpha_{-1}|\tilde{\Omega}\rangle   \\ \hline 
\rule{0mm}{8mm}~~1 & \frac{3}{2} & \mbox{odd}&\fbox{$\psi_{-\frac{1}{2}}\beta_{-\frac{1}{2}}\gamma_{-\frac{1}{2}}|\widetilde{\Omega}\rangle$}\,,\,\fbox{$\psi_{-\frac{3}{2}}|\widetilde{\Omega}\rangle$}\,,\,   c_{-1}\beta_{-\frac{1}{2}}|\widetilde{\Omega}\rangle & \\ 
& && b_{-1}\gamma_{-\frac{1}{2}}|\widetilde{\Omega}\rangle\,\,,
 \,\,G_{-\frac{3}{2}}|\widetilde{\Omega}\rangle \,\,,\,\,\alpha_{-1}\psi_{-\frac{1}{2}}|\widetilde{\Omega}\rangle & \\ \hline
\end{array}
\end{eqnarray*} 
\caption{The states contributing to the potential for $w$ up to level $\frac{3}{2}$. The boxed states 
can be ignored since they are odd under the $\bbbz_{2}$ symmetry which is a combination of the twist and $(\alpha_{-n},\psi_{-n-\frac{1}{2}})\mapsto (-\alpha_{-n},-\psi_{-n-\frac{1}{2}})$.}
\end{table}
Where in Table 2 the boxed states are not invariant 
\cite{0007153,9705038} under the combined $\bbbz_{2}$ twist transformation and 
$
(\psi_{-n-\frac{1}{2}},\alpha_{-n})\mapsto (-\psi_{-n-\frac{1}{2}},-\alpha_{-n})\,.
$
Thus, for example,  the field corresponding to the state $c_{0}\beta_{-\frac{1}{2}}|\widetilde{\Omega}\rangle $
can be set to zero.
\subsubsection{Level $(\frac{1}{2},1)$}
The states contributing at this level and the corresponding vertex operators are
\bea
|\widetilde{\Omega}\rangle \,,\,\,\,\,\,\widehat{T}=\xi \,c\, e^{-\phi}\otimes \sigma_{1}\,;\,\,
\psi_{-\frac{1}{2}}|\widetilde{\Omega}\rangle \,,\,\, \widehat{W}=\xi \,\psi\, c\, e^{-\phi}\otimes I\,.
\eea
Plugging the string field $\widehat{\Phi}=t\,\widehat{T}\,+\,w\,\widehat{W}$ in the superstring action eq(\ref{actionsusy}) and evaluating the terms up to level 1 we get
\bea
g^{2}S=\fracs{{1}}{{4}}\,t^{2}-\fracs{{1}}{{2}}\,t^{4}\,+\,(\frac{1}{2}-\sqrt{2})\,t^{2}w^{2}\,.
\eea
Solving for the equation of motion of $t$ we get
\be
t=0,\,\,\mbox{and}\,\,\,t=t_{\pm}=\pm \frac{1}{2}\sqrt{1-\frac{14}{1+2\sqrt{2}}\,w^{2}}\,.
\ee
Thus we get
\bea
V^{(\frac{1}{2},1)}_{eff}(w)&=&0\,\,\,\mbox{for}\,\,t=0\, \mbox{  ($M$-branch)},\\ \nn
V^{(\frac{1}{2},1)}_{eff}(w)&=&-\frac{2\pi^{2}M}{32}(1-\frac{14}{1+2\sqrt{2}}\,w^{2})^{2}\,\,\,\mbox{for}\,\,\,t=t_{\pm}\, \mbox{  ($V$-branch)}.
\eea
These potentials are shown in the Fig 2 below
\begin{figure}[ht]
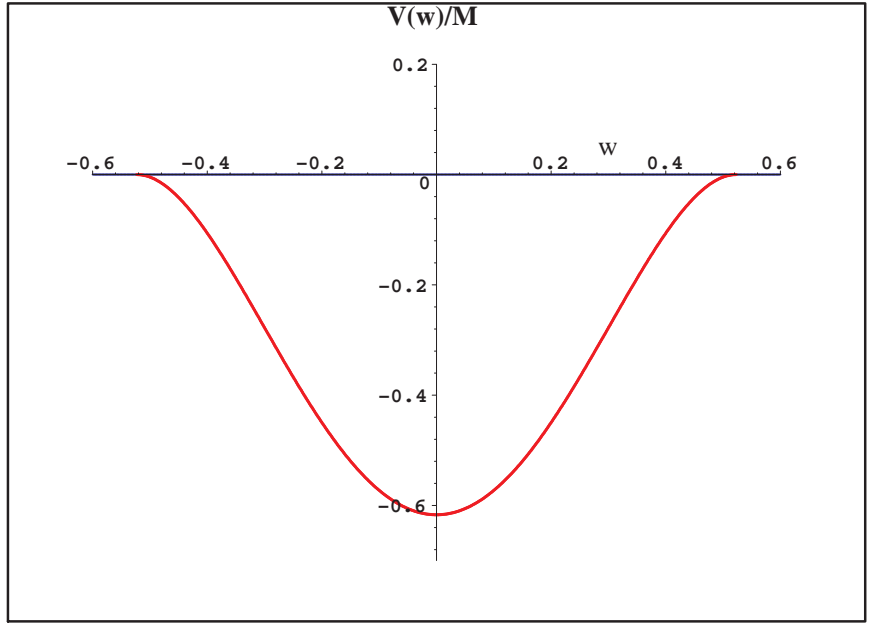

\label{halfone}
\PSbox{halfone.ps hoffset=30 voffset=280 hscale=50 vscale=50 angle=-90}{4.8in}{3.4in}
\caption{The effective potential for $w$ at level $(\frac{1}{2},1)$.} 
\end{figure}
. Note that in the
tachyonic branch the allowed value of $w$ is restricted to the
interval $[-0.52293,+0.52293]$. Also the value of the effective potential 
at $w=0$ in the tachyonic branch is equal to the 
value of tachyon potential at the minimum ($-0.617 M$) calculated in  \cite{0001084}.
\newpage
\subsubsection{Level $(\fracs{{1}}{{2}},2)$}
In this approximation we keep the fields up to level $\fracs{{1}}{{2}}$
and terms in the action up to level $2$. This gives us one more term in the action,
\bea
g^{2}S=\fracs{{1}}{{4}}\,t^{2}\,-\,\fracs{{1}}{{2}}\,t^{4}\,+\,(\frac{1}{2}\sqrt{2})\,t^{2}w^{2}\,-\,\fracs{3}{8}\,w^{4}\,.
\eea
Solving for the  equation of motion of $t$ we get,
\bea
t=0\,,\,\,\,t=t_{\pm}=\pm \frac{1}{2}\sqrt{1-\frac{14}{1+2\sqrt{2}}\,w^{2}}\,.
\eea
In the $M$-branch ( $t=0$ )the effective potential for $w$ is
\bea
V^{(\frac{1}{2},2)}_{eff}(w)&=&(2\pi^{2}M)\frac{3}{8}\,w^{4}\,,\,\,w\in[-\infty,+\infty]\, \mbox{ ($M$-branch)}.
\eea
In the $V$-branch in which $t$ is non-zero we get
\bea\nn
V^{(\frac{1}{2},2)}_{eff}(w)&=&\frac{\pi^{2}M}{16(1+2\sqrt{2})^{2}}\{(48\sqrt{2}-88)\,w^{4}+(56\sqrt{2}+28)\,w^{2}-4\sqrt{2}-9)\,
\mbox{ ($V$-branch)}. 
\eea
and the range of $w$ is the same as it was for the $V$-branch in the $(\frac{1}{2},1)$ approximation.
Fig 3 shows the effective potentials in the $M$ and the $V$-branch.
\begin{figure}[ht]
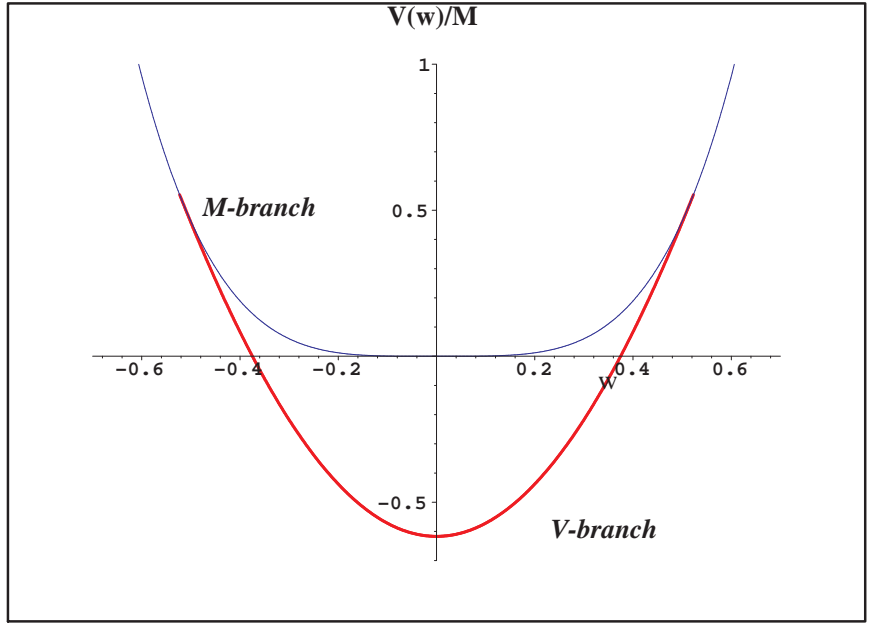

\PSbox{halftwo.ps hoffset=30 voffset=280 hscale=50 vscale=50 angle=-90}{4.8in}{3.4in}
\caption{The effective potential for $w$ at level $(\frac{1}{2},2)$.} 
\end{figure}
\newpage
\subsubsection{Level $(1,2)$}
At this level we have on more state contributing to the potential for $w$,
\bea
\alpha_{-1}|\widetilde{\Omega}\rangle \,,\,\,\,\,\widehat{Y}=i\, \xi\, \partial X\, c\, e^{-\phi}\otimes \sigma_{1}\,.
\eea
The string field is now given by $
\widehat{\Phi}=t\,\widehat{T}\,+\,w\,\widehat{W}\,+\,y\,\widehat{Y}\,.
$
Keeping terms up to level 2 in the action we get,
\bea
g^{2}S=\fracs{{1}}{{4}}\,t^{2}\,-\,\fracs{{1}}{{2}}\,t^{4}\,+\,(\frac{1}{2}-\sqrt{2})\,t^{2}w^{2}\,-\,\fracs{3}{8}\,w^{4}\,-\,kyt-\frac{1}{4}\,y^{2}-\frac{5}{4}\,t^{2}y^{2}\,.
\eea
 We solve for $y$ and $t$ as a function of $w$ by solving for the equations of motion $ \frac{\de S}{\de y}=\frac{\de S}{\de t}=0$
and
get the effective potential for $w$. The effective potential in the two 
branches is shown in Fig 4 below.
 
\begin{figure}[ht]
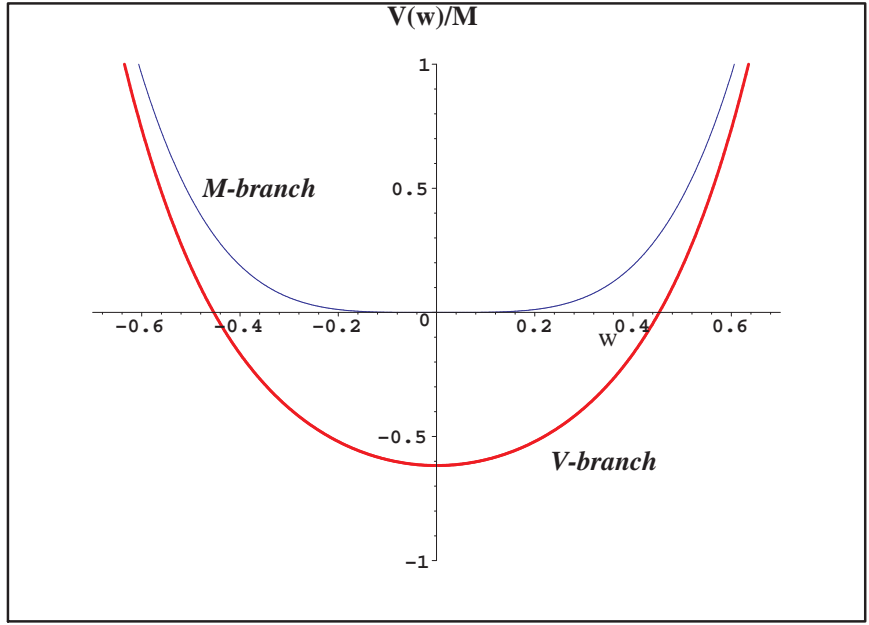

\label{onetwo}
\PSbox{onetwo.ps hoffset=30 voffset=280 hscale=50 vscale=50 angle=-90}{4.8in}{3.4in}
\caption{The effective potential for $w$ at level $(1,2)$.} 
\end{figure}
The effective potential in the $t=0$ branch is the same as the $(\frac{1}{2},2)$ case,
\be
V^{(1,2)}_{eff}(w)=(2\pi^2 M)\, \frac{3}{8}w^{4}\,\,\,\,\,\,\,\mbox{       ($M$-branch)}.
\ee
Although we found an exact analytic expression for the effective potential in the $V$-branch, 
we will not list it here as it is quite complicated. We will just give the first few
terms in its taylor expansion:
\bea \nn
V^{(1,2)}_{eff}(w)=M(-0.61685\,+\,2.31822\,w^2\,+\,2.33284\,w^4\,+4.450471\, w^6\,+\,O(w^8)) \,\,\mbox{ ($V$-branch)\,.}
\eea
\newpage
\subsubsection{Level $(1,3)$}
Including level three terms in the action we obtain one more term as compared to the action at level $(1,2)$,
\bea
g^{2}S&=&\fracs{{1}}{{4}}\,t^{2}\,-\,\fracs{{1}}{{2}}\,t^{4}\,+\,(\frac{1}{2}-\sqrt{2})\,t^{2}w^{2}\,-\,\fracs{3}{8}\,w^{4}\,-\,wyt-\frac{1}{4}\,y^{2}\\ \nn
&-&\frac{5}{4}\,t^{2}y^{2}+(\frac{1}{8}-\frac{1}{2}\sqrt{2})w^{2}y^{2}.
\eea
As before, we can eliminate $y$ and $t$ by using their equations of motion
to obtain the effective potential for
$w$.  In the $M$-branch
 $t=y=0$ and the effective potential for $w$ is
$(2\pi^{2}M)\frac{3}{8}w^{4}$. In the  $V$-branch $t$ and $y$ are
non-zero. In this branch the value of $w$ is restricted to lie in the
range $[-0.7306,0.7306]$. At ends of this interval, the $V$-branch touches the $M$-branch. 
The expression for the effective potential in the $V$-branch is fairly long so we will
just give the first few terms in its taylor expansion:
\bea\nn
V^{(1,3)}_{eff}=M(-0.61685\,+\,2.31822 \,w^2\,+\,4.60253\,w^4\,+\,2.82462\,w^6+O(w^8))\,\,\,\,\mbox{($V$-branch)}\,.
\eea
The graph of the effective potential in
the two branches is shown in figure 5.
\begin{figure}[h]
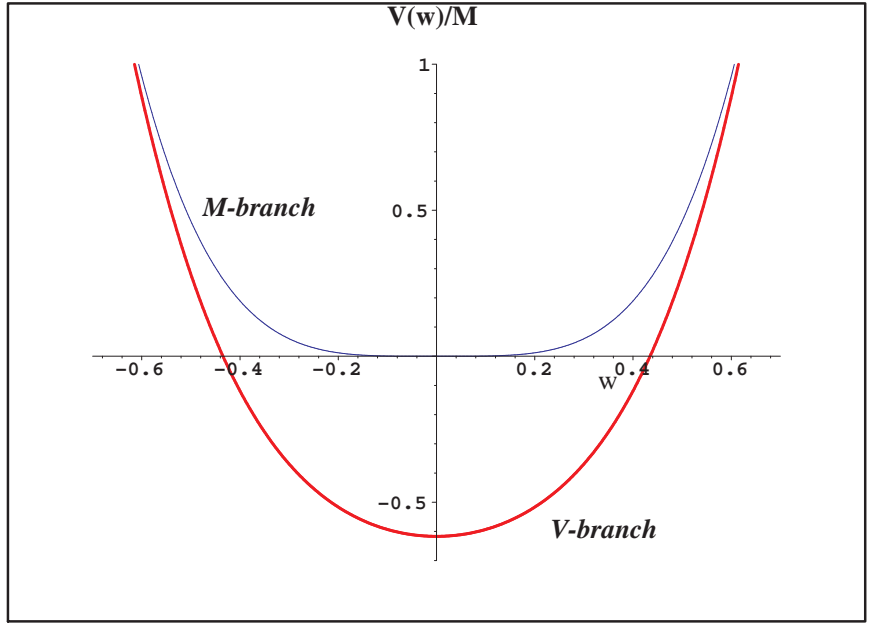

\label{onethreef}
\PSbox{onethree.ps hoffset=30 voffset=280 hscale=50 vscale=50 angle=-90}{4.8in}{3.4in}
\caption{The effective potential for $w$ at level $(1,3)$.} 
\end{figure}
\newpage
\subsubsection{Level $(\frac{3}{2},3)$}
At this level we have four new fields contributing to the potential for $w$,
\bea
\{\,c_{-1}\beta_{-\frac{1}{2}}\,,\,b_{-1}\gamma_{-\frac{1}{2}}\,,\,G_{-\frac{3}{2}}\,,\,\alpha_{-1}\psi_{-\frac{1}{2}}\,\}|\widetilde{\Omega}\rangle 
\eea
with vertex operators
\bea \nn
\widehat{A}=\xi\,\partial\xi\, c \,\partial^{2}c\, e^{-2\phi}\otimes I,\,
\widehat{E}=\xi\,\eta\otimes I,\, \widehat{F}=\xi\, G_{m}\,c \,e^{-\phi}\otimes I,\, \widehat{P}=i\,\xi\, \psi\, \partial X\, c\, e^{-\phi}\otimes I\,.  
\eea
The string field up to this level is given by
\bea
\widehat{\Phi}=t\,\widehat{T}\,+\,w\,\widehat{W}\,+\,y\,\widehat{Y}\,+\,a\,\widehat{A}\,+\,e\,\widehat{E}\,+\,f\,\widehat{F}\,+\,p\,\widehat{P}\,.
\eea
Using this string field and keeping only terms up to level 3 in the action we get, 
\bea
g^{2}S_{0}&=&\frac{1}{4}t^{2}-\frac{1}{2}t^{4}\,,\\ \nn
g^{2}S_{\frac{1}{2}}&=&0\,,\\ \nn
g^{2}S_{1}&=&(\frac{1}{2}-\sqrt{2})t^{2}w^{2}\,,\\ \nn
g^{2}S_{\frac{3}{2}}&=&-wyt+at^{2}+\frac{1}{4}et^{2}+\frac{5}{96}\sqrt{50+22\sqrt{5}}\,et^{4}\,,\\ \nn
g^{2}S_{2}&=&-\frac{3}{8}w^{4}-\frac{1}{4}y^{2}-\frac{5}{4}t^{2}y^{2}\,\\ \nn
g^{2}S_{\frac{5}{2}}&=& \frac{1}{24}(9+5\sqrt{5})\,ew^{2}t^{2}+\frac{4\sqrt{3}}{9}(aw^{2}-fw^{2}-pw^{2})\,,\\ \nn
g^{2}S_{3}&=&-2ae-pf-\frac{5}{18}e^{2}t^{4}-\frac{1}{12}(3+40\sqrt{2})\,aet^{2}+\frac{5}{12}(10\sqrt{2}-1)\,eft^{2}\,\\ \nn
&+&\frac{1}{24}(10\sqrt{2}-1)ept^{2}-\frac{1}{4}(4\sqrt{2}-1)fpt^{2}-5f^{2}-\frac{1}{2}p^{2}-\frac{1}{8}(4\sqrt{2}-1)p^{2}t^{2}\\ \nn
&+&(\frac{1}{\sqrt{2}}-\frac{1}{24})e^{2}t^{2}+(\frac{1}{8}-\frac{1}{2}\sqrt{2})y^{2}w^{2}-\frac{5}{4}(4\sqrt{2}-1)f^{2}t^{2}+\frac{5}{12}(\sqrt{2}-2)eywt\\ \nn
&+&(\frac{3}{4}-\frac{9}{8}\sqrt{2})(fywt+pywt)\,.
\eea
\[
V(w,t,y,a,e,f,p)=-(2\pi^2 M)\,g^2 S
\]
We solve equations of motions for $y,a,e,f,p$ to obtain the vevs of these fields as functions
of $w$ and $t$. Substituting these vevs in $V(w,t,y,a,e,f,p)$, we obtain an effective 
potential $V_{eff}(w,t)$ for $w$ and $t$. This is shown in Fig 6. 
To obtain the effective potential for $w$, we have to solve
the equation $\frac{\de V(w,t)}{\de t}=0$ for $t$ as a function of $w$. There is
alway a solution $t=0$ to this equation. In this branch,  
 the effective potential is given by
\bea
V_{eff}^{(\frac{3}{2},3)}(w)=\frac{17}{216}\,w^{4}\,
\,\,\,\,\,\mbox{$M$-branch}.
\eea
In the $V$-branch $t$ is non-zero and we numerically find the set of points (other than $t=0$)
$(w,t)$ such that $\frac{\de V(w,t)}{\de t}=0$. These points are shown in the Fig 7.
\begin{figure}[p]
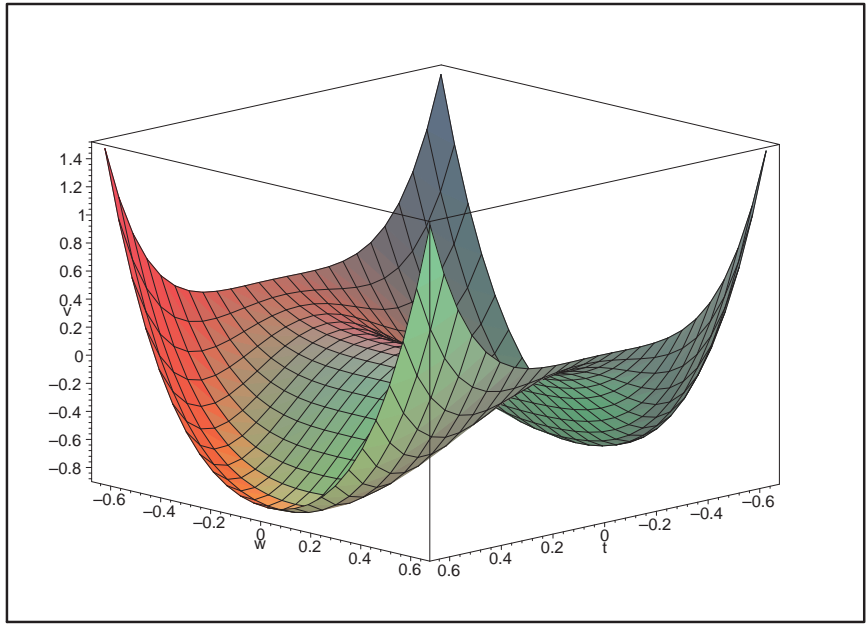

\label{pot3d}
\PSbox{pot3d.ps hoffset=30 voffset=280 hscale=50 vscale=50 angle=-90}{4.8in}{3.4in}
\caption{$\frac{V(w,t)}{M}$ at level $(\frac{3}{2},3)$.} 
\end{figure}
\begin{figure}[p]
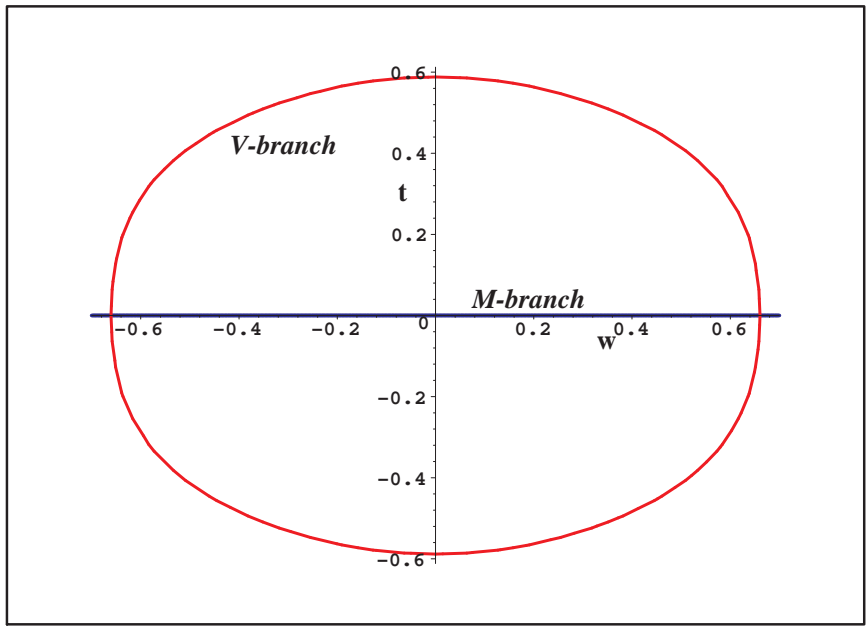

\label{contour}
\PSbox{contour.ps hoffset=30 voffset=280 hscale=50 vscale=50 angle=-90}{4.8in}{3.4in}
\caption{The set of points $(w,t)$ such that $\frac{\de V(w,t)}{\de t}=0$. 
The value of $w$ is restricted to lie in the range $\approx [-0.6617,0.6617]$ and at the end points of the
interval the $V$-branch touches the $M$-branch.} 
\end{figure}
\newpage
From Fig 7 we see that in the $V$-branch, the value of $w$ is restricted to lie
in the range $[-0.6617,0.6617]$. We numerically find the effective
potential for $w$ in the $V$-branch at this level. It is shown in
Fig 8 below. Note that the value of the effective
potential at $w=0$ is $-0.85446$ which is equal to the value of the
tachyon potential at its minimum at this level \cite{0002211}.
\begin{figure}[h]
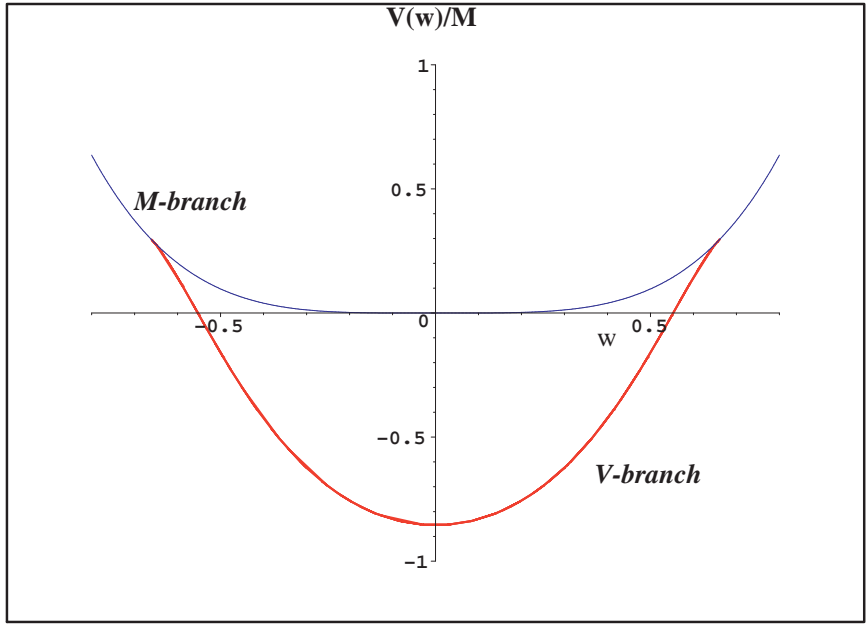

\label{levelthree}
\PSbox{levelthree.ps hoffset=30 voffset=280 hscale=50 vscale=50 angle=-90}{4.8in}{3.4in}
\caption{The effective potential for $w$ at level $(\frac{3}{2},3)$.} 
\end{figure}

\section{Mass of the $U(1)$ gauge field}
The puzzle involving the fate of the $U(1)$ gauge field living on the
non-BPS D-brane, as the D-brane annihilates, has been discussed by
various authors\cite{9911116,9901159,0002223,0008033}. In
\cite{0008033} Taylor calculated the terms quadratic in the gauge
field using level truncation in the bosonic string field theory. He
showed that the eigenvalues of the mass matrix
 approach a non-zero
value as higher and higher levels are included. Using the effective
potential in the $V$-branch determined in the last section we
calculate the mass of the gauge field in the stable tachyonic
vacuum. The mass (squared) is the coefficient of the quaratic term in
the expansion of the effective action in powers of $w$.  The result is given in Table 3.
\begin{table}[ht]
\begin{eqnarray*} 
\begin{array}{||c|c|c||} \hline
\rule{0mm}{5mm}\mbox{Level}         & V(0)           & m^{2}=\fracs{1}{2}V''(0)  \\ \hline \hline
\rule{0mm}{5mm} (\frac{1}{2},1)     &  -0.61685      & 4.51146            \\  \hline
\rule{0mm}{5mm} (\frac{1}{2},2)     &-0.61685        & 4.51146            \\ \hline
\rule{0mm}{5mm} (1,2)               &-0.61685        & 2.31822             \\ \hline
\rule{0mm}{5mm}(1,3)                &-0.61685        & 2.31822             \\ \hline
 \rule{0mm}{5mm} (\frac{3}{2},3)    &-0.85446        & 2.38223 \\ \hline
\end{array}
\end{eqnarray*} 
\caption{The mass of the gauge field in different level approximations. The values 
in the second column are equal to the values, at that level,  of the tachyon potential at its minimum.}
\end{table}
\newpage
It seems that the mass of
the gauge field approaches a non-zero stable value as the level approximation
is improved. However, we will have to include higher
level fields and terms in the action to get more evidence for this fact.

\section{Conclusions}
In this paper we calculated the effective potential for the marginal
deformation parameter corresponding to the constant vev of the $U(1)$ gauge
field in the level truncated superstring field theory. We saw that
for
 the case of BPS D-brane that there are no multiple branches of
the effective
 potential at level $(1,2)$. Therefore their was no
restriction on the range of the marginal deformation parameter
$w$. This was clear from the beginning since
 terms at this level were
at most quadratic in the fields $a,e,f$ and
 $p$ so that equations of
motion are linear in these variables and
 there is a unique vev of
these fields in terms of $w$.  It is possible that if we include level
three terms in the action
 there might be multiple branches of the
effective potential since, in this
 case, we will have terms cubic in
the fields $a,e,f$ and $p$ in the action. For
 example terms like
$f^{3}$, $p^{3}$ and $f^2p$ are non-zero and solving for
 their
equation of motion may give multiple solutions. Although
 determining
the contribution of these cubic terms to the action 
 is relatively
easy, solving for their equations of motion to
 obtain the effective
potential for $w$ seems difficult. However, to find out if the range
of $w$ is restricted in the BPS case and if there are multiple branches
of solutions coming from massive string fields, such a calculation should
be performed.

 For the non-BPS D-brane,  the
existence of two branches of solutions is due to
 to the fact that the tachyon
potential at this level has two extrema
 (up to signs). In the unstable
vacuum corresponding to zero tachyon vev
 ($M$-branch) the effective
potential for $w$ is not restricted and in
 the approximation we
considered is purely quartic with the coefficient
 becoming smaller as
the higher level contributions are
 included.
In the $M$-branch, all GSO($-$) fields are set to zero and the effective
potential is identical to the BPS case (the effective potential
in the BPS case in section 2  is more complicated
than the potential found in the $M$-branch
 of the non-BPS case in section
3; this
is because the level assignments used in the two
 cases are
different). The $V$-branch of the effective potential corresponds to
the
 non-zero tachyon vev and its range is restricted. At the end
point of range
 the $V$-branch merges into the $M$-branch. At $w=0$,
the value of the potential
 in the $V$-branch is the same as the value
of tachyon potential at the minimum
 calculated in
\cite{0001084,0002211}.  
 
 Using the effective potential in the
$V$-branch we also calculate the
 mass of the gauge field in the stable tachyonic
vacuum. Our results seem
 to suggest that the gauge field acquires a
non-zero mass after
 the tachyon condenses. However, to obtain better
evidence of this, the level
 approximation should be pushed further. 

\section*{Acknowledgements}
We thank Barton Zwiebach for useful discussions. This research was supported in part by the US Department of Energy
under contract \#DE-FC02-94ER40818.

\end{document}